\documentclass[10pt, conference, usenames,dvipsnames,svgnames,x11names,tablesquare,numbers]{IEEEtran}

\usepackage[utf8]{inputenc}
\usepackage[T1]{fontenc}
\usepackage{hyphenat} %
\PassOptionsToPackage{hyphens}{url}\usepackage{hyperref}
\usepackage{amsmath,amssymb,amsfonts}
\def\BibTeX{{\rm B\kern-.05em{\sc i\kern-.025em b}\kern-.08em
    T\kern-.1667em\lower.7ex\hbox{E}\kern-.125emX}}
\usepackage[english]{babel}
\usepackage{booktabs}
\usepackage[usenames,dvipsnames,svgnames,x11names,table]{xcolor}
\usepackage{graphicx}
\usepackage{listings}
\usepackage{multirow}
\usepackage{url}
\usepackage{xspace}
\usepackage{ifthen}
\usepackage{caption}
\usepackage{subcaption}
\usepackage{natbib}
\usepackage{enumitem}
\usepackage{textcomp}
\usepackage{adjustbox}
\usepackage{array}
\usepackage{rotating}
\usepackage{footnote}
\usepackage[normalem]{ulem}
\usepackage{pifont} %
\usepackage[abbreviations,xlatin]{foreign}  %
\usepackage[binary-units,per-mode=symbol]{siunitx}
\usepackage{tikz}
\usepackage{collcell}
\usepackage{hhline}
\makesavenoteenv{tabular} 
\makesavenoteenv{table} 
\usepackage{cleveref}
\newlist{inlinelist}{enumerate*}{1}
\setlist*[inlinelist,1]{%
  label=(\roman*),
}
\usepackage{framed}
\usepackage{tcolorbox}
\colorlet{shadecolor}{Azure2}

\hypersetup{
  colorlinks,
  linkcolor={blue!50!black},
  citecolor={blue!70!black},
  urlcolor={blue!70!black}
}

\usetikzlibrary{automata, positioning}

\tcbset{
    colback=Azure2,
    boxrule=0.5pt, 
    boxsep=1pt,
    left=1pt,
    right=1pt,
    top=1pt,
    bottom=1pt,
    sharp corners,
    colframe=white,
}

\date{}
\title{Scrooge Attack:\\Undervolting ARM Processors for Profit}
\IEEEspecialpapernotice{\small{(Practical experience report)}}
\author{\IEEEauthorblockN{Christian Göttel\IEEEauthorrefmark{1}, Konstantinos Parasyris\IEEEauthorrefmark{2}, Osman Unsal\IEEEauthorrefmark{3}, Pascal Felber\IEEEauthorrefmark{1}, Marcelo Pasin\IEEEauthorrefmark{1}, Valerio Schiavoni\IEEEauthorrefmark{1}}
\\
\vspace{-10pt}
\IEEEauthorblockA{
\IEEEauthorrefmark{1}Universit\'e de Neuch\^atel, Switzerland, \texttt{first.last@unine.ch}\\
\IEEEauthorrefmark{2}\textit{Lawrence Livermore National Laboratory}, \texttt{parasyris1@llnl.gov}\\
\IEEEauthorrefmark{3}\textit{Barcelona Supercomputing Center}, \texttt{osman.unsal@bsc.es}\\
}
}

\newcommand\copyrighttext{\footnotesize \textcopyright 2021 IEEE.
Personal use of this material is permitted.
Permission from IEEE must be obtained for all other uses, in any current or future media, including reprinting/republishing this material for advertising or promotional purposes, creating new collective works, for resale or redistribution to servers or lists, or reuse of any copyrighted component of this work in other works.
Presented in the \href{https://srds-conference.org/}{40th IEEE International Symposium on Reliable Distributed Systems (SRDS '21)}.
The final version of this paper is available under DOI: \href{https://doi.org/10.1109/SRDS53918.2021.00027}{10.1109/SRDS53918.2021.00027}
}
\newcommand\copyrightnotice{\begin{tikzpicture}[remember picture,overlay]
\node[anchor=south,yshift=10pt,fill=yellow!20] at (current page.south) {\fbox{\parbox{\dimexpr\textwidth-\fboxsep-\fboxrule\relax}{\copyrighttext}}};
\end{tikzpicture}}

\definecolor{rdylbu1}{HTML}{D73027}
\definecolor{rdylbu2}{HTML}{FC8D59}
\definecolor{rdylbu3}{HTML}{FEE090}
\definecolor{rdylbu4}{HTML}{FFFFBF}
\definecolor{rdylbu5}{HTML}{E0F3F8}
\definecolor{rdylbu6}{HTML}{91BFDB}
\definecolor{rdylbu7}{HTML}{4575B4}

\makeatletter
\newcommand{\ApplyGradient}[1]{%
  \ifdim\dimexpr#1pt>1.10pt
    \hspace{-0.33em}\colorbox{rdylbu1}{}
  \else
    \ifdim\dimexpr#1pt>1.05pt
      \hspace{-0.33em}\colorbox{rdylbu2}{}
    \else
      \ifdim\dimexpr#1pt>1.02pt
        \hspace{-0.33em}\colorbox{rdylbu3}{}
      \else
        \ifdim\dimexpr#1pt>0.98pt
          \hspace{-0.33em}\colorbox{rdylbu4}{}
        \else
          \ifdim\dimexpr#1pt>0.95pt
            \hspace{-0.33em}\colorbox{rdylbu5}{}
          \else
            \ifdim\dimexpr#1pt>0.90pt
              \hspace{-0.33em}\colorbox{rdylbu6}{}
            \else
              \hspace{-0.33em}\colorbox{rdylbu7}{}
            \fi
          \fi
        \fi
      \fi
    \fi
  \fi
}

\newcommand{\ApplyGradientBW}[1]{%
  \ifdim\dimexpr#1pt>1.10pt
    \edef\x{\noexpand\cellcolor{black!0}}\x\textcolor{black}{\tiny\bf\sffamily #1}%
  \else
    \ifdim\dimexpr#1pt>1.05pt
      \edef\x{\noexpand\cellcolor{black!10}}\x\textcolor{black}{\tiny\bf\sffamily #1}%
    \else
      \ifdim\dimexpr#1pt>1.00pt
        \edef\x{\noexpand\cellcolor{black!20}}\x\textcolor{black}{\tiny\bf\sffamily #1}%
      \else
        \ifdim\dimexpr#1pt=1.00pt
          \edef\x{\noexpand\cellcolor{black!40}}\x\textcolor{black}{\tiny\it\sffamily #1}%
        \else
          \ifdim\dimexpr#1pt>0.95pt
            \edef\x{\noexpand\cellcolor{black!60}}\x\textcolor{white}{\tiny\sffamily #1}%
          \else
            \ifdim\dimexpr#1pt>0.90pt
              \edef\x{\noexpand\cellcolor{black!70}}\x\textcolor{white}{\tiny\sffamily #1}%
            \else
              \edef\x{\noexpand\cellcolor{black!80}}\x\textcolor{white}{\tiny\sffamily #1}%
            \fi
          \fi
        \fi
      \fi
    \fi
  \fi
}
\makeatother

\newcolumntype{A}{%
  >{\collectcell\ApplyGradient}%
  c%
  <{\endcollectcell}%
}

\newcolumntype{R}[2]{%
  >{\adjustbox{angle=#1,lap=\width-(#2)}\bgroup}%
  l%
  <{\egroup}%
}
\newcolumntype{G}{>{\collectcell\ApplyGradientBW}{r}<{\endcollectcell}}

\newcommand*\rot{\multicolumn{1}{R{45}{1em}}}

\begin{document}

\maketitle
\thispagestyle{plain} %
\pagestyle{plain}     %
\copyrightnotice
\begin{abstract}
Latest ARM processors are approaching the computational power of x86 architectures while consuming much less energy.
Consequently, supply follows demand with Amazon EC2, Equinix Metal and Microsoft Azure offering ARM-based instances, while Oracle Cloud Infrastructure is about to add such support.
We expect this trend to continue, with an increasing number of cloud providers offering ARM-based cloud instances.

ARM processors are more energy-efficient leading to substantial electricity savings for cloud providers.
However, a malicious cloud provider could intentionally reduce the CPU voltage to further lower its costs.
Running applications malfunction when the undervolting goes below critical thresholds.
By avoiding critical voltage regions, a cloud provider can run undervolted instances in a stealthy manner.

This practical experience report describes a novel attack scenario: an attack launched by the cloud provider against its users to aggressively reduce the processor voltage for saving energy to the last penny.
We call it the Scrooge Attack and show how it could be executed using ARM-based computing instances.
We mimic ARM-based cloud instances by deploying our own ARM-based devices using different generations of Raspberry Pi.
Using realistic and synthetic workloads, we demonstrate to which degree of aggressiveness the attack is relevant.
The attack is unnoticeable by our detection method up to an offset of \SI{-50}{\milli\volt}.
We show that the attack may even remain completely stealthy for certain workloads.
Finally, we propose a set of client-based detection methods that can identify undervolted instances.
We support experimental reproducibility and provide instructions to reproduce our results.
\end{abstract}
\section{Introduction}
\label{sec:intro}
Cloud providers continuosly upgrade their commercial offerings to adapt to market and customer needs.
While the vast majority of them offer computing instances based on x86 processors, the availability of ARM-based cloud instances is quickly expanding.
ARM processors are increasing their market share of server-grade machines~\cite{ampere:emag,ampere:altra,huawei:kunpeng,nvidia:grace,aws-graviton,aws:graviton2,cavium:thunderx,fujitsu:hotchips}, thanks to additional energy and performance improvements.
Before ARM announced its Neoverse~\cite{arm:neoverse} microarchitecture, there were no server-grade ARM processors to license.
Companies had to customize application-grade ARM processor designs for their server-grade platforms~\cite{aws-graviton,cavium:thunderx}.
Recent ARM server-grade processors~\cite{ampere:emag,huawei:kunpeng,cavium:thunderx,fujitsu:hotchips} are based on custom-developed ARMv8 microarchitectures.
For example, Amazon~\cite{aws-graviton} deploys ARM-based processors currently shipped in off-the-shelf ARM hardware.
Their AWS Graviton processor is essentially a more powerful quad-Raspberry Pi 4B~\cite{raspberry:pis}.
Scaleway offered instances based on custom-made ARM SoCs with servers smaller than a business card~\cite{scaleway:preview}.
\autoref{tab:arm} summarizes a subset of available server-grade ARM processors,  supported instruction set architectures (ISA), and providers deploying this hardware.
Several generations of ARM processors~\cite{ampere:emag,ampere:altra,aws-graviton,aws:graviton2,cavium:thunderx} are currently available across cloud providers.
ARM processors also started reaching into the supercomputing market segment.
We expect an increasing availability of ARM Neoverse processors and future server-grade ARM instances to close the performance gap to x86. %
\begin{table}[t]
  \centering
  \caption{List of server-grade and mimicking ARM processors with their supported ISA. `\textbf{*}': used in our evaluation (see \S\ref{sec:eval}).\label{tab:arm}}
  \setlength{\aboverulesep}{0pt}
  \setlength{\belowrulesep}{0pt}
  \rowcolors{1}{gray!10}{gray!0}
  \begin{tabular}{>{\kern-\tabcolsep}lll<{\kern-\tabcolsep}}
    \toprule\rowcolor{gray!25}
    \multicolumn{1}{c}{\textbf{Processor}} & \multicolumn{1}{c}{\textbf{ISA}} & \multicolumn{1}{c}{\textbf{Cloud provider}}
    \\ \midrule
    Ampere Altra & ARMv8.2+ & Equinix, Oracle \\ %
    Ampere eMAG 8180 & ARMv8 & Equinix \\ %
    AWS Graviton & ARMv8 & AWS \\
    AWS Graviton 2 & ARMv8.2 & AWS \\
    Fujitsu A64FX & ARMv8.2 & - \\ %
    Huawei Kunpeng 920 & ARMv8.2 & - \\
    Marvell ThunderX & ARMv8 & Equinix \\ %
    Marvell ThunderX2 & ARMv8.1 & Microsoft Azure \\
    NVIDIA Grace & TBA & - \\
    \midrule
    Broadcom BCM2837(B0)\textbf{*} & ARMv8 & - \\
    Broadcom BCM2711\textbf{*} & ARMv8 & - \\ %
    \bottomrule
  \end{tabular}
\end{table}
On the one hand processor manufacturers specify conservative voltage margins due to process variation~\cite{papadimitriou2017harnessing}.
On the other hand processors offer different power management mechanisms to adjust frequencies and voltages.
While marginal energy savings on a single device appear unimportant, it is of importance at scale, especially since power savings of the cloud infrastructure accumulate for each CPU.
The energy footprint of a single execution step (\ie one single instruction on a processor) is fairly independent of the CPU frequency but dependent on the CPU voltage~\cite{hicss95}.
Decreasing the CPU voltage below the nominal value to conserve power is called undervolting\footnote{Notice that Dynamic Voltage and Frequency Scaling (DVFS) differs from undervolting by decreasing frequency as well as voltage.}. %
Besides energy savings, undervolting directly influences core temperature and can also reduce core aging~\cite{vanSanten2016avs}. %
Undervolting, however, incurs the risk of introducing soft~\cite{parasyris2018framework} and hard-errors related to timing violations~\cite{papadimitriou2020limits}. %
These types of errors can be mitigated by carefully analyzing the guardband of processors~\cite{koutsovasilis2020dynamic}. %
In this practical experience report, we consider a scenario where processors supporting a cloud infrastructure are undervolted by an excessively economic and malicious cloud provider (a \emph{scrooge} \S\ref{subsec:cloudprovider}) to profit from additional electricity bill savings, while cloud users (from here on referred to as users) observe similar performance. %
Unfortunately, undervolting cannot be applied arbitrarily.
In fact, it comes at the cost of processor reliability when the supplied voltage is insufficient to drive the processor's frequency.
We believe this is a risk that malicious cloud providers are willing to take. %
For users, undervolting opens up a new attack vector against their cloud applications (see our threat model in \S\ref{sec:model}).
The main research questions we address in this work are:
\begin{tcolorbox}
    \textit{\textbf{RQ1:} What is necessary for a malicious cloud provider in order to pull off a stealthy undervolting strategy?}
\end{tcolorbox}
\begin{tcolorbox}
    \textit{\textbf{RQ2:} Does a cloud user have the ability to uncover such an undervolting strategy?}
\end{tcolorbox}
To answer those questions, we need to lay the foundation to better understand consequences of (arbitrary) undervolting, both from the cloud provider and client perspective. 
In fact, depending on supply voltage, frequency, load, and temperature of the CPU, execution steps can yield erroneous computations.
While recent attacks~\cite{plundervolt,kenjar2020v0ltpwn} have demonstrated how undervolting can be effectively exploited to gain access to sensitive information, we deal with a different threat model: the infrastructure is undervolted on purpose by a powerful attacker (\ie, the cloud provider), at the risk of exposing hard-to-detect unreliable computing instances for users.
Without physical access to instances, nor being able to directly manipulate the supply voltage or frequency, a user's options remain limited.
Nevertheless, a user can adjust the processor's load and operating performance points (\S\ref{ssec:pwrmgmt}) to influence its heat dissipation.
In order to operate under full load, the processor has to be set to the highest operating performance point, which implies the highest frequency and supply voltage setting.
Consequently, undervolted processors present higher probability for erroneous computations to occur because they are unable to maintain high frequencies.
This probability is further increased by the propagation delay due to high operating temperature.
If erroneous computations result in faults, one can observe application crashes, or kernel panics, leading to cloud instance unavailability. 
While service level agreements (SLA)~\cite{patel2009service} typically cover such scenarios, a malicious provider might try to balance its actions to only yield erroneous computations not resulting in faults, basically overcoming SLA protections. %
For this reason, we designed a non-selective fault injection method for detecting the scrooge attack.
The sole purpose of the detection method is to yield intentional application crashes or kernel panics on undervolted instances such that the user is covered by the SLA.
While interesting, we consider cloud providers or users exploiting undervolting to leak sensitive information~\cite{tang2017clkscrew,chen2021voltpillager} to be out of scope of this work.

\begin{figure}[!t]
  \centering
  \includegraphics[width=.82\linewidth]{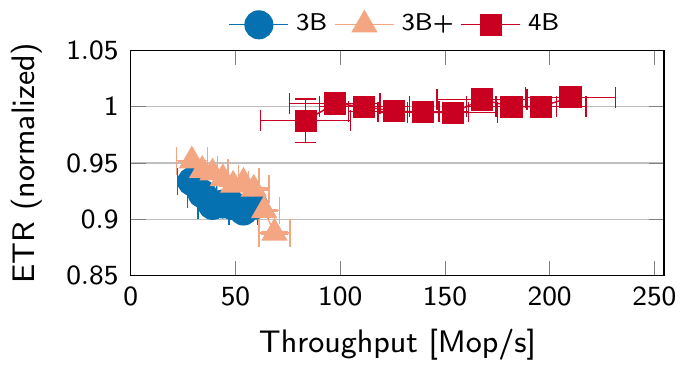}
  \caption{Normalized energy to throughput ratio (ETR) with absolute errors for undervolted Raspberry Pi model B platforms operating at maximum throughput\label{fig:motivation}}
\end{figure}
Interestingly, ARM-based Raspberry Pis have already been collocated in cloud data centers~\cite{raspberry:pcextreme}.
With the intent to reproduce and study the dynamics of such deployments (and, to a smaller scale, mimic AWS using ARM nodes), we first study the effects 
of undervolting on three different ARM processors, focusing on energy savings. %
\autoref{fig:motivation} shows different normalized energy to throughput ratios (ETR)~\cite{hicss95} obtained with ARM Cortex-A processors for the three latest Raspberry Pi models (3B, 3B+, and 4B~\cite{raspberry:pis}) at their lowest operational undervolting setting (\SI{-75}{\milli\volt} for 3B and 3B+, and \SI{-15}{\milli\volt} for 4B) compared to nominal voltage (\ie \SI{0}{\milli\volt}, no undervolting). %
As shown, undervolting directly influences energy spent per operation, without negatively affecting throughput.
Lower normalized ETR values indicate higher energy efficiency for a given throughput.
On average across different throughput values we achieved by undervolting \SIrange{5}{13}{\percent} better energy efficiency on the 3B and 3B+ and \SIrange{0}{3}{\percent} on the 4B.
In essence, these results suggest that a cloud provider can indeed undervolt ARM-based instances, without directly compromising the observed performance.

Our contributions are as follows:
\begin{itemize}
  \item We describe a novel attack scenario based on undervolting by a scrooge cloud provider to lower energy costs.
  \item We demonstrate how cloud users can with a certain probability detect this novel scrooge attack.
  \item We provide a temperature-based guardband analysis to narrow down the operation voltage range of an ARM-based processor (\S\ref{subsec:analysis}).
  \item We describe how our analysis can be used to automatically identify undervolted instances (\S\ref{subsec:investigation})%
  \item We present potential energy gains of undervolting systems using a reliability benchmark (\S\ref{subsec:reliability}).
    In general gains can reach up to \SI{37}{\percent}. %
\end{itemize}

This practical experience report is organized as follows.
\Cref{sec:background} provides background on the low-level mechanisms used to undervolt a processor and the Raspberry Pi platform as well as the associated side-effects.
Our threat model is given in \Cref{sec:model}.
We overview our detection method in  \Cref{sec:impl}.
Our in-depth experimental evaluation is presented in  \Cref{sec:eval}.
We discuss and review related work in \Cref{sec:disc} and \Cref{sec:rw}, before concluding in \Cref{sec:conc}.
\section{Background}
\label{sec:background}
This section defines more precisely a few concepts related to power management (\S\ref{ssec:pwrmgmt}), \ie frequency and voltage scaling and associated techniques such as Dynamic Voltage and Frequency Scaling (DVFS) and Adaptive Voltage Scaling (AVS).
In \S\ref{ssec:reliability} we explain the relation between such techniques and how they affect the overall reliability of a system.

\subsection{ARM in data centers}
Collocation offers allow users to either ship or buy Raspberry Pis in order to deploy lightweight workloads on this low-energy hardware and thus free up resources on high-energy x86 hardware.
Furthermore, Raspberry Pis are the size of credit cards and have much lower cooling demands, which allows hosting a large number of units in a single rack~\cite{scaleway:preview}.
Such off-the-shelf hardware setups allow for large-scale node deployments as needed in data processing or cloud computing workloads.
While off-the-shelf hardware typically lacks in performance and storage capability, its energy consumption remains comparable to server hardware. %

\begin{figure}[!t]
  \centering
  \includegraphics[width=\linewidth]{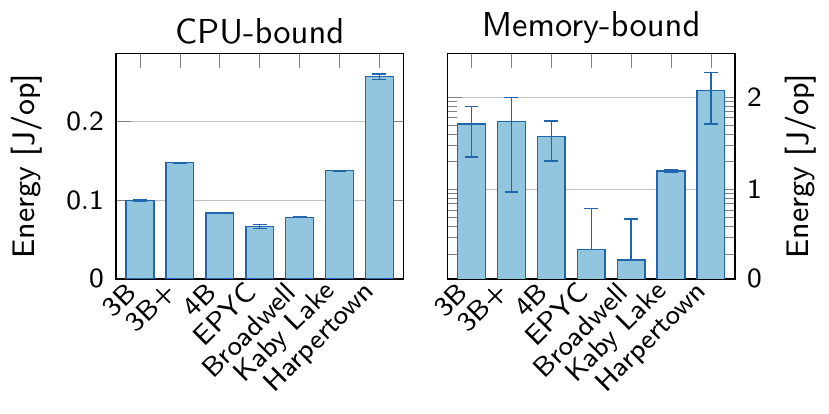}
  \caption{Energy comparison of off-the-shelf and server-garde devices on CPU-bound and memory-bound workloads.}\label{fig:ecomp} %
\end{figure}

\autoref{fig:ecomp} compares the energy consumption of ARM-based off-the-shelf hardware (\ie, three different Raspberry Pi models) against server-grade hardware using different x86 architectures.
We run a cryptographic (CPU-bound) and a memory allocation (Memory-bound) stressor while measuring the entire device power consumption.
The x86 processors used were an AMD \emph{EPYC} and three different Intel Xeon processor generations, \ie,  \emph{Broadwell},  \emph{Kaby Lake} and \emph{Harpertown}.
This is a direct comparison of the execution of two distinct binaries of the same source code on two different architectures based on a common metric (\si{\joule\per op}).
Despite a large difference in power consumption between ARM and x86 hardware, their energy consumption per operation is comparable.
For a detailed discussion on energy metrics with comparison of different architectures we refer to~\cite{hicss95}.
We observe no major difference for CPU-bound operations between different architectures~\cite{blem2013power}.
However, memory-bound operations on off-the-shelf hardware have higher energy consumption.
In the case of the Raspberry Pi models, these are due to cache size and memory transfer rate.
Nevertheless, off-the-shelf hardware achieves lower energy consumption for both operations compared to older server-grade x86 hardware, \ie Harpertown.
These results indicate that replacing old x86 hardware with recent off-the-shelf ARM-based nodes in data centers will result in energy savings.

\subsection{Power management}\label{ssec:pwrmgmt}

The power dissipated by an integrated circuit depends on static power (leakage current) and dynamic power (switching power).
Since about 2005~\cite{esmaeilzadeh2011darksilicon} the power dissipation contribution of dynamic power has become much higher than static power.
Nowadays, with the decreased transistor size and lowered threshold voltages, static power is becoming more and more important~\cite{hotpower10}.
In the following, we outline techniques to reduce dynamic power.

\textbf{Frequency scaling} regulates (dynamically) the frequency of an integrated circuit in order to change performance, conserve power or reduce the amount of heat dissipation.
Reducing the frequency at a constant voltage is called \emph{underclocking} or throttling, while increasing the frequency is called \emph{overclocking}.
The dynamic power dissipated by an integrated circuit over a period of time is given by $P = C V^2 f$, where $C$ is the capacitance, $V$ is the voltage, and $f$ is the frequency.
Thus, increasing the frequency results in a higher power consumption and operating temperature.

\textbf{Voltage scaling} is an open loop system, in which the voltage of an integrated circuit is regulated (dynamically) based on an external setting.
Increasing or decreasing the voltage while keeping the frequency constant is called \emph{overvolting} and \emph{undervolting}, respectively.
Regulating the voltage enables increasing the frequency or conserving power of an integrated circuit, a particularly useful aspect especially for battery-powered devices.
Changing the voltage influences the rate at which capacitances can be charged and discharged.
Thus voltage determines the speed and frequency at which an integrated circuit can be operated.
Modern operating systems do not provide direct support to adjust a processor's voltage individually.
The processor's voltage is either regulated by model-specific registers~\cite{undervolt:msr} or through firmware.

\textbf{DVFS} is the simultaneous software-controlled regulation of voltage and frequency scaling of an integrated circuit.
Depending on the process variation (variation of integrated circuits when fabricated) ARM system on a chip (SoC) manufacturers specify a set of operating performance points (OPPs) under worst case conditions.
These OPPs are pairs of clock frequencies and voltages under which the integrated circuit is operational with a sufficiently large margin while taking into account thermal conditions.
In Linux the CPUFreq kernel driver~\cite{linux:cpufreq} will chose a set of OPPs based on a specified governor.
DVFS has been extensively studied~\cite{hotpower10,weiser1994energy,10.5555/2643634.2643655} to accelerate multi-threaded applications.
x86 manufacturers use their own DVFS implementations~\cite{intel:speedstep,amd:coolnquiet,amd:powernow}.

\textbf{AVS}~\cite{nielsen1994low}
is a closed loop system where the voltage is regulated based on its process variation, aging and a feedback loop of sensor data.
A hardware monitor or software backed by sensor data determines if the changes made to the system are sufficient or if additional changes are necessary.
AVS requires support from both the processor and the power regulators, in order to adjust the voltage accordingly.
The Raspberry Pi models B used in this report are equipped with an AVS system.

\subsection{Raspberry Pi}\label{ssec:rapi}
The Raspberry Pi's firmware is configured at boot time by a text file containing property-value pairs.
For example, the frequency and voltage can be set in this configuration file.
A particularity is that voltages can only be set to a nominal offset in steps of \SI{25}{\milli\volt}.
These offset steps are referred to as overvoltage levels in the Raspberry Pi firmware configuration.
This configuration file is then parsed by the firmware.
While this undervolting configuration is specific to the Raspberry Pi, other hardware can more easily be undervolted dynamically at runtime.
Notice that the requested CPU frequency in the operating system can deviate from the actual frequency regulated by the firmware.
This is in particular the case if the device is throttled for reaching the thermal hard limit at \SI{85}{\degreeCelsius}.
Additionally, the 3B+ has a soft limit temperature at \SI{60}{\degreeCelsius} that will throttle the CPU frequency and voltage.

\subsection{Reliability}\label{ssec:reliability}

There are several approaches to determine a processor's reliability in an undervolted operating regime.
Known benchmarks (\eg, SPEC CPU2006~\cite{henning2006spec}, PARSEC~\cite{bienia08characterization}, \etc) are still used~\cite{koutsovasilis2020dynamic,zu2015adaptive,tan20183}. %
Recently, new specialized power viruses~\cite{MICRO12,CAL17} have been proposed to maximize power consumption and voltage noise.
Even small proof-of-concept programs are sufficient for fault detection under voltage and/or frequency scaling~\cite{plundervolt,tang2017clkscrew}.
Finally, such programs can also be used to characterize the guardband of a system~\cite{tang2017clkscrew,DSN20}.

In this practical experience report we distinguish between three regions with respect to the guardband: safe, critical, and failure.
A safe region has a sufficiently high voltage margin, such that erroneous computations or transient faults cannot occur.
The critical region designates a small voltage band in which the processor occasionally experiences erroneous results or transient faults.
Inside the failure region it is impossible to boot the operating system either because the voltage cannot support the processor's frequency or because erroneous computations and transient faults lead to kernel crashes or panics.

Undervolted instances become unavailable in case a transient fault leads to an instance crash. %
From our perspective, current SLAs cover single instances that have crashed because of undervolted hardware, provided users can sufficiently support these claims.
The situation is trickier with multiple instances.
Deployed instances would have to crash simultaneously, yet process variation plays into the cloud provider's hands.
These crashes are non-deterministic and, therefore, process variation helps obfuscating the undervolted setup.
Only simultaneous crashes satisfy today's cloud provider restrictions in order for users to be covered by the SLA. %
\section{Threat model}
\label{sec:model}
In this section we discuss our threat model.
In particular, we intend to clarify: \emph{(1)} which techniques can a malicious cloud provider use to hide an undervolted processor from an unsuspecting user, and \emph{(2)} which are the methods for a curious user to reveal an undervolted processor?
Notice that we validate these methods on a specific hardware configuration (\ie, Raspberry Pi boards using Broadcom BCM2837/BCM2711 processors), but the discussion holds for other platforms relying on similar voltage regulation mechanisms.

\subsection{The scrooge cloud provider}
\label{subsec:cloudprovider}
We assume the cloud provider has full access to the physical infrastructure and can connect remotely to the physical machines~\cite{aws:encryption}. %
Furthermore, the cloud provider purposefully undervolts its ARM-based hardware to benefit from additional savings. %
Firmware configurations can be hidden from users for malicious or security purposes.
By maliciously intercepting any voltage reading requests (see \S\ref{subsec:attack}), the cloud provider ensures that the undervolted state of the cloud infrastructure remains oblivious to users.
A cloud provider must find the sweet spot~\cite{koutsovasilis2020dynamic} for the undervolt configuration in or near the critical region to provide sufficiently stable instances. %

\subsection{The curious cloud user}
\label{subsec:clouduser}

The curious cloud user is suspicious of the cloud provider and intends to uncover its potentially obfuscated activity.
Instances of the cloud provider can exclusively be accessed remotely by the user.
The only way for a user to detect an undervolted processor is by querying the firmware, normally using a specific executable command file for that.
By reading values from the firmware and comparing them to values in the boot configuration file, a user can detect an undervolted processor.
If results of firmware queries can be forged, it becomes difficult for a user to uncover the scrooge cloud provider.
A confidential and tamper-proof message exchange with the firmware is essential to detect an undervolted processor.

A user can suspect an undervolted processor to operate in the critical region in case of kernel warnings or kernel panics appearing during the system boot or while the system is running, despite these being generic kernel warnings rather than specific ones. %
In particular if the booting time is longer than expected, then this might hint at a failed boot attempt where the kernel crashed.
Most systems have a kernel log that can be consulted by the system administrator.
However, a cloud provider can tamper with those kernel logs, and the system utilities are outside the trusted computing base.

\subsection{The scrooge attack}
\label{subsec:attack}
\begin{figure}[!t]
  \centering
  \includegraphics[scale=0.65]{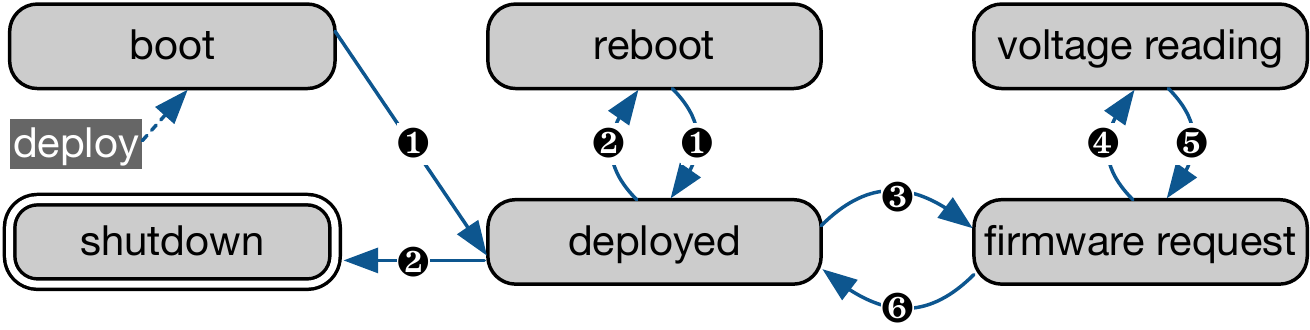}
  \caption{State machine with cloud provider actions to obfuscate undervolted machine configuration.}\label{fig:sm}
\end{figure}

The scrooge cloud provider makes undervolted ARM instances available to users.
These undervolted instances should be indistinguishable from nominal voltage instances.
This includes configuration, firmware, and tools querying CPU voltages.
Thus, the undervolt configuration needs to be exchanged for a nominal configuration and any CPU voltage reading request needs to be intercepted.
\autoref{fig:sm} shows different actions the cloud provider has to perform during an instance lifecycle in order to hide the undervolt configuration.
When a user boots such an instance, the cloud provider must ensure that the undervolt configuration is loaded by the firmware on the machine the instance is running on.
However, this undervolt configuration should not be accessible once the user is connected to the instance.
The undervolt configuration has to be swapped for the nominal configuration (Fig.\labelcref{fig:sm}-\ding{182}).
Depending on the configuration mechanism, the file system that was booted may be different from the file system the user finds after booting.
This includes firmware, operating system kernel, binaries, \etc.
A hidden or obfuscated system service could perform this task while the operating system is booting.
An even stealthier approach involves a trusted operating system~\cite{optee:homepage} or auxiliary devices~\cite{aws:nitro} which exchange the configurations before the operating system is booted.
Therefore, without proper system attestation, there is no guarantee about the authenticity of the system users believe they have booted.
During reboot or shutdown of the machine configurations might have to be swapped back (Fig.\labelcref{fig:sm}-\ding{183}) again.

Any CPU voltage reading request needs to be intercepted and substituted by a plausible nominal voltage value.
This will typically involve a kernel driver that will handle the communication with the firmware or accessing model-specific registers.
The request can then be intercepted directly in the user space tool or the kernel driver (Fig.\labelcref{fig:sm}-\ding{184}).
From the kernel driver the request is forwarded (Fig.\labelcref{fig:sm}-\ding{185}) and the actual undervolted CPU voltage value is returned to the kernel driver (Fig.\labelcref{fig:sm}-\ding{186}).
The kernel driver then substitutes this value by a some nominal voltage value, \eg by adding the undervolt offset to the value.
The alleged nominal voltage value is then returned to the user (Fig.\labelcref{fig:sm}-\ding{187}).
A more costly but stealthier variant involves the trusted operating system, to which the cloud provider could delegate voltage reading requests instead of a kernel driver. %

If users are allowed to deploy their own kernels, then the cloud provider needs a different approach.
Voltage reading request can no longer be intercepted in kernel space.
Instead, the cloud provider needs to use the hypervisor to intercept CPU voltage requests and substitute them similarly to the kernel driver approach.

In our threat model we assume that the cloud provider will make use of these mechanisms and obfuscate as much as possible the undervolted state of the infrastructure from users, a practical effort with significant benefits.
Without access to the firmware configuration nor any untampered message exchange with the CPU voltage regulating mechanism, a user can never be sure to obtain a genuine voltage reading.

\subsection{Relevance of discussed techniques}
While a scrooge cloud provider has powerful mechanisms in place to hide its undervolted instances, the curious user can still expose this misbehaviour.
For instance, the processor's frequency and package temperature are viable options to test for undervolted conditions.
The techniques presented in \Cref{sec:eval} demonstrate to which extent users can deploy applications stressing aforementioned options on instances and how accurately conclusions can be drawn.

\section{Scrooge attack detection}
\label{sec:impl}

This section describes the user's detection method as well as under which conditions the detection method works and where difficulties may arise.
Furthermore, we mention potential concerns of cloud providers.

We assume that users cannot trust any firmware or system reading on instances.
As such, users have no reference to any parameters for adjusting the detection method to the attack.
Users can for this reason make use of simple CPU-bound programs that will put the processor under maximum load while monitoring for faults.
Inspired by \cite{plundervolt} we propose implementing an arithmetic computation (\ie multiplication) for which we can validate the result.
First we generate two random numbers which are then multiplied until the instance crashes while alternating the position of multiplier and multiplicand.
Murdock \etal have observed, that the position of the multiplier and the multiplicand can lead to a faulting instruction.
After each multiplication the result is compared to the original result.
While the processor operates at maximum load it will run at the highest frequency and dissipate heat which will raise its temperature.
Under these conditions we achieve the highest probability to inject faults related to timing violations.
Depending on the complexity of the RISC circuitry in the ARM processor, certain instructions are more likely to fault then others.
To this end, the detection method might not inject faults in its own computation (due to its simple nature), but more likely in other processes.
This behavior is favorable, as it allows to run the detection method until the instance itself becomes unavailable due to multiple critical faults in system relevant processes.
Thus, we detect an undervolted cloud instance using the detection method by %
gradually failing processes to crash the instance and make it unavailable.

The detection method depends strongly on how aggressively machines are undervolted and the cooling system employed by the cloud provider.
The less a machine is undervolted, the higher the temperature needs to be raised by the detection method to fault processes and \viceversa.
A good cooling system is a lesser problem than a weakly undervolted machine.
With a good cooling system the detection method requires a longer time to raise the processor's temperature.
On the one hand implementing a soft limit temperature throttle in order to prevent this detection method is not an ideal solution.
Users are less inclined to pay for a service which underperforms compared to alternative services. %
On the other hand weakly undervolting machines defies the scrooge cloud provider's original idea of minimizing the electricity bill.

The cloud provider's options are limited to completely prevent the detection method from unveiling the scrooge attack.
Even the powerful setup of the cloud provider to tamper with CPU voltage readings is not sufficient denying the detection method.
The scrooge attack has the disadvantage that detection methods have a
simple design, but it has the advantage that proving the undervolt state without the firmware is difficult.
\section{Evaluation}
\label{sec:eval}
In this section we explore the behavior of Raspberry Pi processors under different nominal and undervolted setups.
The information gained from these experiments allows quantifying the attack parameters and determining the type of processes to use for the detection method.
Then, we derive the probability at which our detection method can successfully uncover the attack.
We begin by describing our experimental setup to undervolt Raspberry Pis before evaluating the firmware's throttling behavior when reaching the soft limit and limit temperatures.
The temperature-based guardband analysis allows detecting the critical region of the device and defines the margin for an undervolt setup.
Faults that occurred during the guardband analysis are analyzed to describe the fault injection of the detection method.
Finally, we measure the energy efficiency of the undervolted hardware with a reliability benchmark.
The dataset gathered for this evaluation is publicly available at \url{https://github.com/ChrisG55/Scrooge-Attack}.

\begin{figure*}[t]
  \centering
  \begin{subfigure}[t]{0.315\textwidth}
    \includegraphics[width=\linewidth]{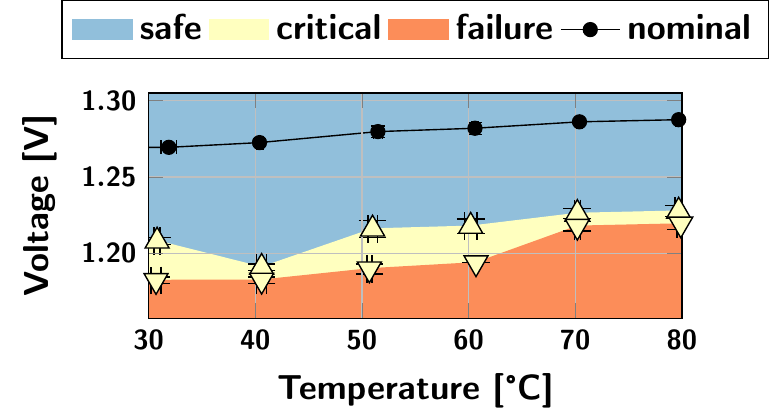}
    \caption{Raspberry Pi 3B}
  \end{subfigure}%
  \begin{subfigure}[t]{0.315\textwidth}
    \includegraphics[width=\linewidth]{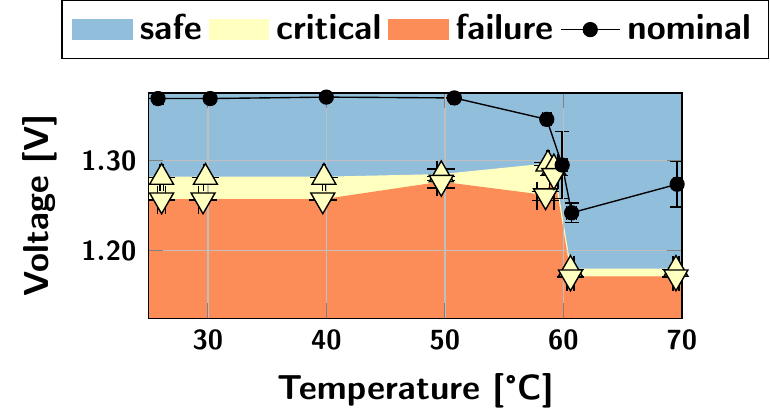}
    \caption{Raspberry Pi 3B+}
  \end{subfigure}%
  \begin{subfigure}[t]{0.2965\textwidth}
    \includegraphics[width=\linewidth]{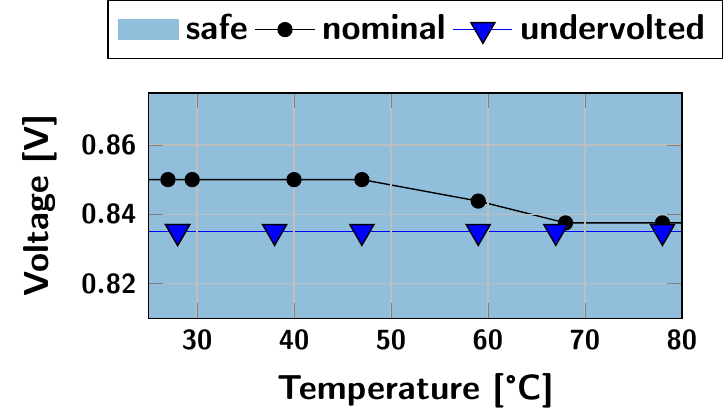}
    \caption{Raspberry Pi 4B}
  \end{subfigure}
  \caption{Temperature-dependent guardband measurements with absolute errors of latest Raspberry Pi models B. Triangles indicate lower (\protect\rotatebox[origin=c]{180}{$\triangle$}) and upper ($\triangle$) frontier measurements while bullets ($\bullet$) indicate nominal measurements.\label{fig:undervolting}}
\end{figure*}

\subsection{Experimental settings}
We use the three Raspberry Pi models 3B, 3B+, and 4B, while booting from the same SD card a Raspbian Buster distribution (\url{https://github.com/raspberrypi/linux}). 
All units rely on recent firmware releases (since June 2020).
To simulate a realistic cloud scenario we take all measurements in an air-conditioned room at \SI[separate-uncertainty]{24(1)}{\degreeCelsius} and connect the Raspberry Pis to Ethernet and run the SSH daemon.
The Raspberry Pis are monitored over UART from an auxiliary machine.
No other peripherals are connected to the Pis in order to minimize any interference.
Both the Raspberry Pi's and the auxiliary machine's clocks are synchronized using NTP, to easily correlate the power consumption logs recorded on the auxiliary machine to the benchmark running on the Raspberry Pi.
The power consumption of the Raspberry Pi is recorded by an Alciom PowerSpy2~\cite{alciom:powerspy} over bluetooth.
The entire experimental setup can be fully automated for a manageable effort using a network-capable power strip.
\begin{table}[!t]
  \centering
  \caption{Soft limit (SL) firmware throttling on the 3B+}
  \label{tab:softlimit}
  \setlength{\aboverulesep}{0pt}
  \setlength{\belowrulesep}{0pt}
  \rowcolors{1}{gray!10}{gray!0}
  \begin{tabular}{>{\kern-\tabcolsep}crrrr<{\kern-\tabcolsep}}
    \toprule\rowcolor{gray!25}
    \multicolumn{1}{c}{\textbf{OV level}} & \multicolumn{1}{c}{\textbf{V$_{\mathrm{arm}}$ [\si{\volt}]}} & \multicolumn{1}{c}{\textbf{f$_{\mathrm{arm}}$ [\si{\mega\hertz}]}} & \multicolumn{1}{c}{\textbf{V$_{\mathrm{arm}}^{\mathrm{SL}}$ [\si{\volt}]}} & \multicolumn{1}{c}{\textbf{f$_{\mathrm{arm}}^{\mathrm{SL}}$ [\si{\mega\hertz}]}} \\
    \midrule
    0 & \num{1.3750} & \num{1400} & \num{1.2688} & \num{1200} \\
    \rowcolor{gray!10}
    -1 & \num{1.3500} & \num{1400} & \num{1.2375} & \num{1200} \\
    \rowcolor{gray!0}
    -2 & \num{1.3188} & \num{1400} & \num{1.2125} & \num{1200} \\
    \rowcolor{gray!10}
    -3 & \num{1.2938} & \num{1400} & \num{1.1875} & \num{1200} \\
    \bottomrule
  \end{tabular}
\end{table}

\subsection{Soft limit temperature throttling}
\label{sec:softlim}

We start by evaluating the firmware behavior when reaching the soft limit temperature while running under the CPUFreq \emph{performance} governor.
Understanding the throttling behavior helps evaluating the viability of possible mitigation techniques by the cloud provider.
The Raspberry Pi documentation mentions that frequency and voltage of the SoC are reduced to decrease heat dissipation but without indicating by how much.
The Raspberry Pi 3B+ is the only model with a soft limit temperature programmed into its firmware, therefore, other models are not included in the overvoltage level to OPP mapping reported in \Cref{tab:softlimit}.
The values indicate that the ARM CPU frequency f$_\mathrm{arm}$ is reduced by \SI{200}{\mega\hertz} and the CPU voltage V$_\mathrm{arm}$ is lowered by about \SI{106}{\milli\volt} (four levels).
The voltage stepping of \SI{25}{\milli\volt} remains the same with two exceptions from nominal level $-1\to -2$ with \SI{-31.2}{\milli\volt} and from soft limit level $0\to -1$ with \SI{-31.3}{\milli\volt}.

\begin{table}[!t]
  \centering
  \caption{Limit temperature (L) throttling on the 3B and 4B}
  \label{tab:hardlimit}
  \setlength{\aboverulesep}{0pt}
  \setlength{\belowrulesep}{0pt}
  \rowcolors{1}{gray!10}{gray!0}
  \begin{tabular}{>{\kern-\tabcolsep}crrcrrr<{\kern-\tabcolsep}}
    \toprule\rowcolor{gray!25}
    \multicolumn{1}{c}{\textbf{Model}} & \multicolumn{1}{c}{\textbf{V$_{\mathrm{arm}}^{\mathrm{L}}$ [\si{\volt}]}} & \multicolumn{1}{c}{\textbf{f$_{\mathrm{arm}}^{\mathrm{L}}$ [\si{\mega\hertz}]}} & \multicolumn{1}{c}{\textbf{f$_{\mathrm{core}}^{\mathrm{L}}$ [\si{\mega\hertz}]}} \\
    \midrule
    3B & \num{1.2813} & $\{1034,1087,1141,1195,1200\}$ & $\{400\}$ \\
    \rowcolor{gray!10}
    4B & \num{0.8500} & $\{1000,1500\}$ & $\{333,500\}$ \\
    \bottomrule
  \end{tabular}
\end{table}

\subsection{Limit temperature throttling}
Next, we evaluate the firmware behavior when reaching the limit temperature while running under the CPUFreq \emph{performance} governor.
At the limit temperature, the firmware will throttle the processor to prevent thermal runaway. %
Notice that model 3B+ is not included here, as it is taking too much time reaching the limit temperature while already being throttled for going beyond the soft limit temperature.
Neither the 3B nor the 4B reduce the voltage when reaching the limit temperature as shown in \Cref{tab:hardlimit}.
However, both models reduce their frequency.
For every voltage in the scale, there is a range of frequency values that do not crash the processor.
The sets of frequencies indicated in \Cref{tab:hardlimit} are frequencies downstepped to by the firmware until thermal limits are satisfied when being throttled.
The 3B is reducing its ARM CPU frequency f$^\mathrm{L}_\mathrm{arm}$ in steps of about \SI{54}{\mega\hertz} (except for the first step) while the 4B significantly reduces its frequency by \SI{500}{\mega\hertz}.
In addition the 4B also reduces its GPU frequency f$_{\mathrm{core}}^{\mathrm{L}}$ by \SI{167}{\mega\hertz}.
We find that reaching the limit temperature will reduce the load put on the processor by the detection mechanism and reduce its temperature which lead to a lower fault injection rate.

\subsection{Temperature-based guardband analysis}
\label{subsec:analysis}
The temperature-based guardband analysis helps identifying voltage margins of the Raspberry Pi models.
While this analysis supports the cloud provider in selecting an undervolt offset, its core principle can also be exploited by users to uncover the scrooge attack.
This benchmark consists of three stages: 1) booting the operating system while undervolted before 2) adjusting the SoC's temperature either actively or passively and 3) running a billion iterations of the multiplication benchmark described in \S\ref{sec:impl} as a single-threaded process. %
We set the CPUFreq governor to \emph{performance} right before starting the multiplication benchmark.
This will guarantee that the multiplication benchmark is started at a well defined temperature and that it runs at a constant, maximum frequency.
After each benchmark execution we reduce the ARM CPU voltage level in the configuration and reboot the system.
This process is repeated until the system no longer boots because the supply voltage has gone below threshold or because the system has crashed.

The results of this analysis are shown in \autoref{fig:undervolting}.
All Raspberry Pi models keep a sufficient margin with their nominal voltage (connected black bullets) configuration to the critical region.
Further undervolting of the ARM CPU into the critical region results in occasionally failing processes. %
Undervolting the ARM CPU beyond threshold voltage makes it impossible to boot the hardware.
Our multiplication benchmark, that verifies the correct operation of the ARM CPU, never detected an incorrect result.
We explain this characteristic of the multiplication benchmark, which is purely based on arithmetic operations, by not being on a timing-critical path to force an incorrect operation of the ARM CPU.

We can also see that the undervolting depends directly on the SoC's temperature.
For instance on the Raspberry Pi 3B we clearly observe slightly rising regions, which result from small adjustments made by the AVS system. %
This is mainly due to the resistivity of the circuitry that increases with temperature. %

The Raspberry Pi 4B can only be undervolted once to level \num{-1} at \SI{-15}{\milli\volt} due to missing overclocking~\footnote{\url{https://www.raspberrypi.org/documentation/configuration/config-txt/overclocking.md} Last accessed on 2021-07-13} support in the firmware.
This undervolt limit is indicated by the blue line.
However, we observe some basic overheating protection mechanism that slightly lowers nominal voltage by \SI{-12.5}{\milli\volt} in the range of \SIrange{50}{70}{\degreeCelsius}.

\begin{table*}[t]
	\renewcommand{\arraystretch}{1} %
	\setlength{\fboxsep}{0pt} %
	\setlength{\tabcolsep}{2pt} %
	
  \centering
  \caption{\textsc{stress-ng} ETR heat map indicating the relative energy efficiency for an undervolted setup compared to a nominal setup. The darker the shade, the more energy-efficient the stressor ran.\label{tab:reliability}}
  \begin{tabular}{clc*{27}{|G}|}
    \rot{Cooling} &
    \rot{Model} &
    \rot{Undervolt} &
    \rot{aio} &
    \rot{atomic} &
    \rot{bsearch} &
    \rot{clock} &
    \rot{fork} &
    \rot{futex} &
    \rot{get} &
    \rot{hrtimers} &
    \rot{hsearch} &
    \rot{icache} &
    \rot{judy} &
    \rot{kcmp} &
    \rot{kill} &
    \rot{lsearch} &
    \rot{membarrier} &
    \rot{mergesort} &
    \rot{msg} &
    \rot{pipe} &
    \rot{poll} &
    \rot{sem} &
    \rot{sigsegv} &
    \rot{sysfs} &
    \rot{timer} &
    \rot{tsearch} &
    \rot{urandom} &
    \rot{vm-rw} &
    \rot{wcs}
    \\
    \toprule
    \hhline{~~~---------------------------}
    \parbox[t]{2mm}{\multirow{3}{*}{\rotatebox[origin=c]{90}{active}}}
    & 3B  & \SI{-75}{\milli\volt} & 0.94 & 0.95 & 0.96 & 0.95 & 0.92 & 1.02 & 1.03 & 0.90 & 0.95 & 0.95 & 0.99 & 0.93 & 0.93 & 0.94 & 1.02 & 0.94 & 0.91 & 0.96 & 0.95 & 0.93 & 0.94 & 0.91 & 0.94 & 0.99 & 0.95 & 0.96 & 0.94 \\
    \hhline{~~~---------------------------}
    & 3B+ & \SI{-75}{\milli\volt} & 0.89 & 0.94 & 0.93 & 0.93 & 0.87 & 0.99 & 0.94 & 1.00 & 0.95 & 0.94 & 1.01 & 0.93 & 0.96 & 0.94 & 0.97 & 0.94 & 0.95 & 0.92 & 0.95 & 0.93 & 0.93 & 0.92 & 0.94 & 0.95 & 0.95 & 0.97 & 0.94 \\
    \hhline{~~~---------------------------}
    & 4B  & \SI{-15}{\milli\volt} & 1.01 & 0.99 & 1.02 & 0.99 & 1.02 & 0.98 & 1.00 & 1.06 & 1.00 & 0.98 & 0.96 & 1.00 & 1.04 & 0.99 & 0.96 & 1.00 & 0.97 & 0.70 & 0.98 & 0.98 & 0.99 & 1.00 & 0.97 & 0.91 & 0.98 & 1.00 & 0.99 \\
    \hhline{~~~---------------------------}
    \midrule
    \hhline{~~~---------------------------}
    \parbox[t]{2mm}{\multirow{3}{*}{\rotatebox[origin=c]{90}{passive}}}
    & 3B  & \SI{-75}{\milli\volt} & 0.88 & 0.95 & 0.92 & 0.93 & 0.91 & 0.66 & 0.76 & 0.63 & 0.94 & 0.94 & 0.97 & 0.93 & 0.94 & 0.94 & 0.93 & 0.93 & 0.92 & 1.03 & 0.93 & 0.95 & 0.92 & 0.95 & 0.92 & 0.96 & 0.94 & 0.96 & 0.93 \\
    \hhline{~~~---------------------------}
    & 3B+ & \SI{-75}{\milli\volt} & 0.95 & 0.95 & 0.96 & 0.95 & 0.98 & 0.97 & 0.99 & 0.80 & 0.95 & 0.95 & 0.94 & 0.95 & 0.95 & 0.95 & 0.98 & 0.95 & 0.95 & 0.96 & 0.94 & 0.95 & 0.94 & 0.91 & 0.95 & 0.97 & 0.96 & 0.97 & 0.95 \\
    \hhline{~~~---------------------------}
    & 4B  & \SI{-15}{\milli\volt} & 1.00 & 0.97 & 1.02 & 0.99 & 1.01 & 0.84 & 1.00 & 1.10 & 0.99 & 1.03 & 0.99 & 0.98 & 1.00 & 1.00 & 0.97 & 1.00 & 1.03 & 1.01 & 1.00 & 1.05 & 1.04 & 0.87 & 1.00 & 0.91 & 0.99 & 0.97 & 0.99 \\
    \hhline{~~~---------------------------}
    \bottomrule 
  \end{tabular}
\end{table*}

\begin{figure}[t]
  \centering
  \includegraphics[width=.8\linewidth]{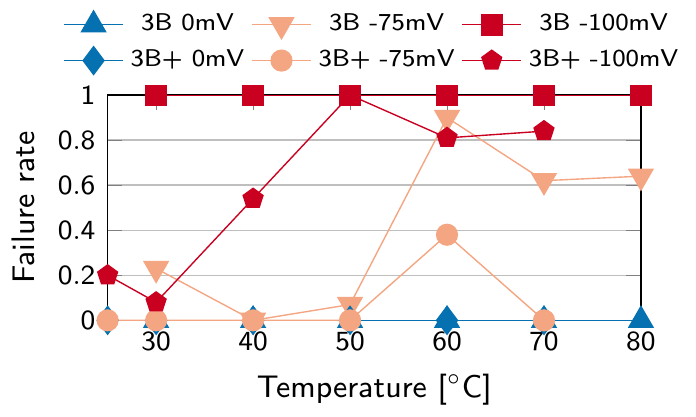}
  \caption{Temperature-dependent failure rate of Raspberry Pi models at different undervolt levels.\label{fig:fail}}
\end{figure}

\begin{figure*}[!t]
  \centering
  \begin{subfigure}[t]{0.2465\textwidth}
    \includegraphics[width=\linewidth]{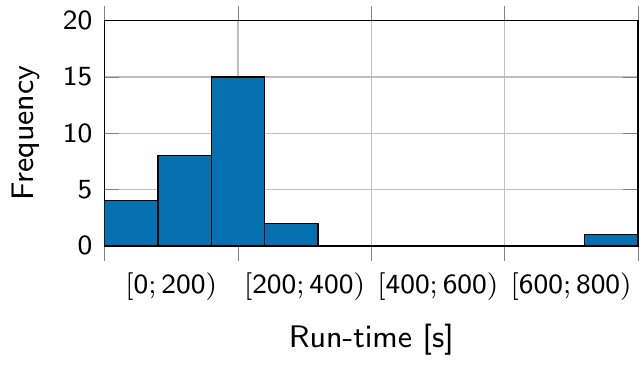}
    \caption{Bare-metal 3B run-time}
  \end{subfigure}%
  \begin{subfigure}[t]{0.2465\textwidth}
    \includegraphics[width=\linewidth]{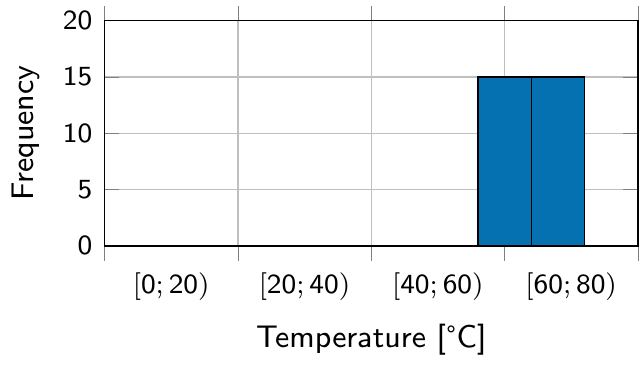}
    \caption{Bare-metal 3B temperature}
  \end{subfigure}%
  \begin{subfigure}[t]{0.2465\textwidth}
    \includegraphics[width=\linewidth]{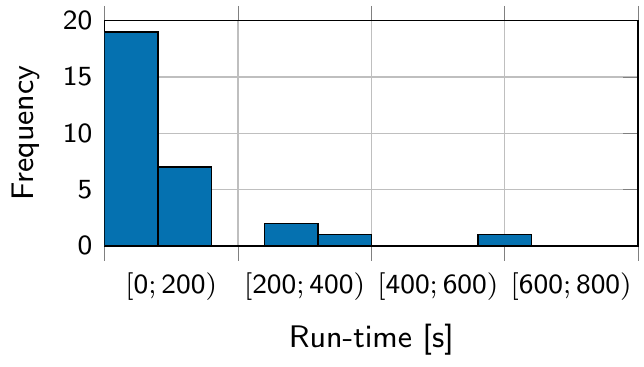}
    \caption{Kubernetes 3B run-time}
  \end{subfigure}%
  \begin{subfigure}[t]{0.2465\textwidth}
    \includegraphics[width=\linewidth]{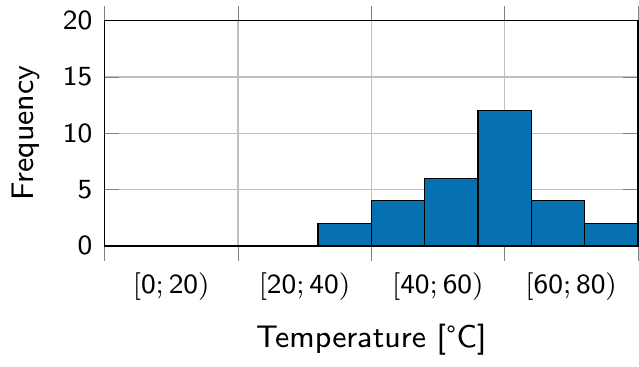}
    \caption{Kubernetes 3B temperature}
  \end{subfigure}\\
  \begin{subfigure}[t]{0.2465\textwidth}
    \includegraphics[width=\linewidth]{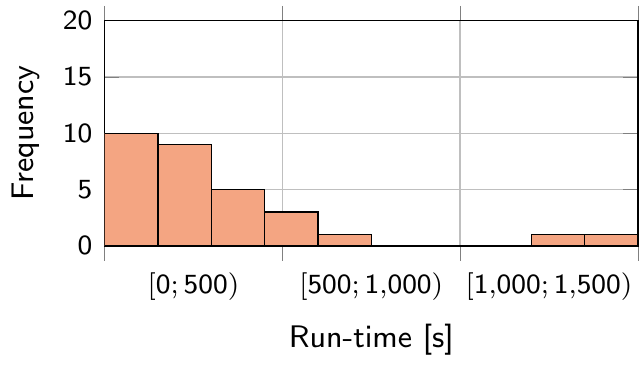}
    \caption{Bare-metal 3B+ run-time}
  \end{subfigure}%
  \begin{subfigure}[t]{0.2465\textwidth}
    \includegraphics[width=\linewidth]{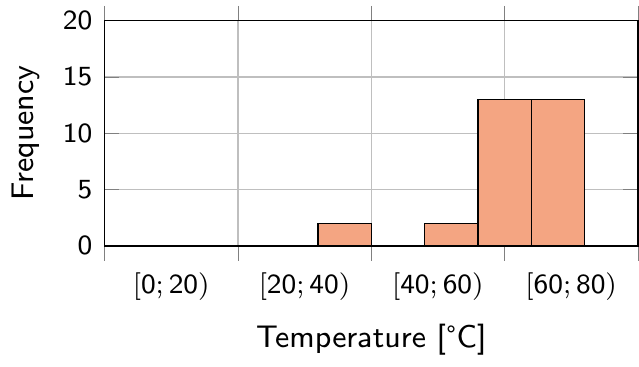}
    \caption{Bare-metal 3B+ temperature}
  \end{subfigure}%
  \begin{subfigure}[t]{0.2465\textwidth}
    \includegraphics[width=\linewidth]{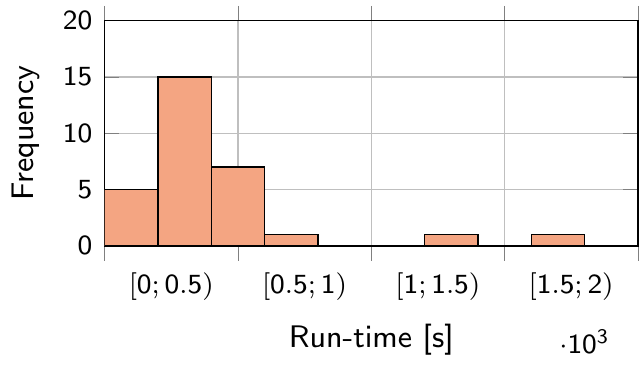}
    \caption{Kubernetes 3B+ run-time}
  \end{subfigure}%
  \begin{subfigure}[t]{0.2465\textwidth}
    \includegraphics[width=\linewidth]{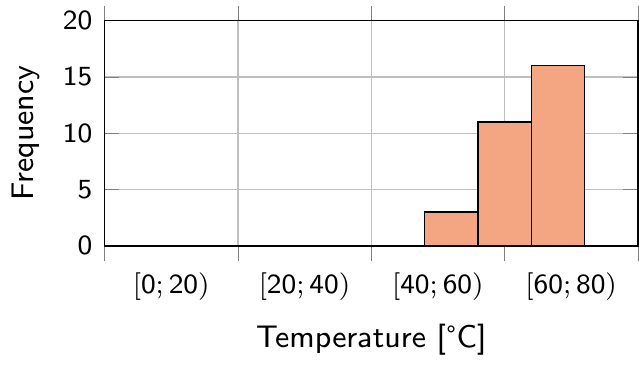}
    \caption{Kubernetes 3B+ temperature}
  \end{subfigure}
  \caption{Run-time and temperature histograms of bare-metal and container instances\label{fig:param}}
\end{figure*}

\subsection{Implications on the detection method success rate}
\label{subsec:investigation}
In order to produce \Cref{fig:undervolting} and to fully characterize safe undervolt parameters we (as cloud provider) ran overall \num{741} guardband analyses.
In \num{265} of those \num{741} guardband analyses the system had experienced a failure or crashed as a result of operating in the critical region.
The ARM architecture provides the Exception Syndrome Register~\cite{arm:reference}, which the operating system can consult to diagnose the type of exception generated by a process.
Among the \num{265} failed or crashed runs we identified \num{407} process failures of the following \num{5} types: \emph{(1)} paging requests (\SI{46.4}{\percent}), \emph{(2)} freeze during boot (\SI{26.7}{\percent}), \emph{(3)} NULL pointer dereferences (\SI{20.3}{\percent}), \emph{(4)} read from unreadable memory (\SI{5.4}{\percent}), and \emph{(5)} write to read-only memory (\SI{0.9}{\percent}).
These types of failures, with the exception of \emph{(2)}, usually generate a kernel oops.
A kernel oops happens when the operating system kernel detects an incorrect behavior of a process and can possibly resume execution of the system.
In some cases execution cannot be resumed because of system dependencies or unavailable system resources as a result of the failing process.
The kernel will raise a panic and halt the system if a kernel oops occurs in an interrupt handler.
If possible, the operating system kernel will log this kernel oops diagnose in the system log.

With the information from the \num{265} failed or crashed guardband analyses we derived a temperature-dependent failure rate for the 3B and 3B+ in \autoref{fig:fail}.
Notice that the 4B is not included, as it's firmware does not provide undervolting support and we could not provoke any failures in this system.
Our analysis indicates that at \SI{60}{\degreeCelsius} the detection method has the highest probability with a \SI{40}{\percent} chance on the 3B+ respectively a \SI{90}{\percent} chance on the 3B to provoke a failure in a system operating in the critical region.
Compare with the results in \Cref{subsec:parameters} from the point of view of a cloud user running the detection method.
With an even more aggressive undervolting at \SI{-100}{\milli\volt}, failures can already be reliably provoked at \SI{40}{\degreeCelsius} on both devices.
These failures were provoked in at least \num{33} different processes (\SI{34}{\percent} user, \SI{15}{\percent} kernel processes and \SI{51}{\percent} unknown processes) of which the multiplication benchmark is never among the known failed processes.
For the 3B+ the failure rate is dropping at \SI{70}{\degreeCelsius} due to the temperature soft-limit throttling in the firmware to bring the system back into the safe region.
At nominal voltage and in the safe region no failures could be provoked in any system during temperature-based guardband analysis execution.

\subsection{``Reliability'' and energy efficiency}
\label{subsec:reliability}
A common option to test the reliability of the selected undervolt parameters for our systems is to run benchmarks.
Despite stressing the systems for several hours, we (as cloud provider) could not provoke any failures in these systems using \textsc{stress-ng}.
Consequently, as cloud provider we would assume, that our chosen undervolt parameters keep the systems in the safe region as we covered successfully a large set of stressors with our benchmark.
However, this assumption on our system's reliability turns out to be deceptive.
These stressors and other benchmarks are not optimized to push all components of these processors to their limits.
For example, the collection of statistics in a stressor's main loop can introduce sufficient overhead for not being able to test a component's limits.
Therefore, given the circumstances, an undervolted processor can suddenly be operating in the critical region.
In our scenario, these circumstances can be changed by raising the heat dissipation of the processor to increase its temperature. 
The detection method is based on this exact principle, where a cloud user tries to trip the cloud instance processor into the critical region by putting the processor under heavy load and raising its temperature.
The interesting aspect of this detection method is that there is no specific requirement on the program putting the processor under heavy load.
Instead, the detection method exploits the nature of all other processes running simultaneously on the system by eventually failing.
While undervolting improves energy efficiency of processors in general, we believe that processors should also be equipped with features improving the reliability while operating under such critical conditions, \ie error correcting codes.
We show in heat map \autoref{tab:reliability} the energy efficiency for all three Raspberry Pi models based on the ETR ratio of the undervolted to the nominal setup.
For the measurements we use two cooling setups: active and passive cooling.
As there are many cooling solutions and because cooling solutions differ among cloud providers, our measurements do not include the energy consumed by our cooling system.
Whether these additional gains outweigh the cost of the active cooling system remains an open but also interesting question.
We ran up to 169 stressors sequentially of which 27 are shown in the heat map.
Each stressor was configured with a timeout of \SI{60}{\second}.

Our results show that when the Raspberry Pi's are actively cooled, we can achieve higher energy efficiency.
It is even possible to undervolt the device further.
For example, with the 3B+ we were able to undervolt up to \SI{-100}{\milli\volt} and in some rare instances also run the benchmark successfully.
One particular observation is made on the 3B that uses an older thermal design.
On the 3B passive cooling results in higher energy efficiency compared to active cooling.
18 of the 27 stressors were more energy efficient with passive cooling than active cooling on the 3B with an average of \SI{-4}{\percent} energy efficiency per stressor.
Occasionally we also observe better energy efficiency in the passive cooling setup than in the active cooling setup on the 3B+ (\eg hrtimers with \SI{-20}{\percent} and judy with \SI{-7}{\percent}) and 4B (\eg futex with \SI{-14}{\percent} and sysfs with \SI{-13}{\percent}).
We noticed that some stressors have a large variance in the number of operations.
Hence, if these stressors achieve a higher than average number of operations during the measurement, their energy efficiency improves proportionally.
With the 4B we notice only minor improvements in energy efficiency.
Again, this is due to the lack of overclocking support but also because of the lower core supply voltage compared to the other models.
For the 3B+ and 4B we can say that a setup with active cooling results in larger energy gains compared to a passive cooling setup.
On average across the 27 stressors \SI{5}{\percent} / \SI{9}{\percent} (active/passive) were saved on the 3B, \SI{6}{\percent} / \SI{6}{\percent} were saved on the 3B+, and \SI{2}{\percent} / \SI{1}{\percent} were saved on the 4B.
The highest energy efficiency observed on the 3B was \SI{-10}{\percent} / \SI{-37}{\percent} on the hrtimers stressor.
On the 3B+ \SI{-13}{\percent} / \SI{-20}{\percent} were saved on the fork / hrtimers stressor.
Finally, the \SI{-30}{\percent} / \SI{-16}{\percent} were saved on the 4B with the hrtimers / futex stressor.

\subsection{Detection method parameters}
\label{subsec:parameters}
In this subsection we quantify the detection method parameters (\ie run-time and temperature) based on undervolted bare-metal and container instances.
Deploying virtual machines on the Raspberry Pi is impracticable and were therefore not included in our evaluation.
To run containers on the Raspberry Pi we deployed a small Kubernetes cluster.
The detection method is deployed by the cloud user on instances using SSH.

\autoref{fig:param} shows histograms with crashes on bare-metal and container instances deployed on the 3B and 3B+.
We show the run-time of our detection method and the temperature at which instances crashed.
Our observations made with the temperature-based guardband analysis in \autoref{subsec:investigation} are confirmed by the temperature histograms.
The run-time strongly depends on the processor's capability to heat up to a certain temperature and is therefore not an ideal parameter.
We observe clear differences between the thermal designs of the two models.
For the 3B our detection method requires about \SI{175}{\second} / \SI{30}{\second} (bare-metal / container) to reach \SI{62}{\degreeCelsius} to crash bare-metal or container instances.
On the 3B+ we require about \SI{145}{\second} / \SI{250}{\second} to reach \SI{62}{\degreeCelsius} to crash bare-metal or container instances.
Interestingly, container instances crash on the 3B earlier than bare-metal instances.
We assume the computing requirements from the container environment and the thermal design work in favor of the detection method.

These histograms assist a cloud user to narrow down the exact parameters for tripping a processor into the critical region and successfully crash instances.
In order to achieve a simultaneous crash of multiple instances, the cloud user has to adjust the timing of the detection method's deployment.
The broad distribution of the run-time in the histograms suggests that the program used in the detection method for putting the processor under load needs to synchronize with other instances to throttle the load in order to achieve a simultaneous crash.
\section{Discussion}
\label{sec:disc}
From our evaluation we conclude that the detection method is best used in combination with other processes such as in \textsc{stress-ng}.
The user even has the option to scale the number of threads in the detection method to adjust the crash time of an instance as well as the injection rate.
A simple CPU-bound program like the multiplication benchmark turns out to be ideal for injecting faults in an undervolted setup.
The advantage of such a simple CPU-bound program is that it is unlikely to inject faults during its own execution and can run until a kernel panic while raising heat dissipation.
In terms of energy efficiency we observed that by undervolting the cloud provider can save on average \SI{5}{\percent} and up to \SI{37}{\percent} for specific workloads on ARM processors.
\begin{tcolorbox}
    \textit{\textbf{RA1:} as shown by our extensive experimental evaluation, in order to pull off a stealthy undervolting strategy, a malicious cloud provider must exchange any firmware configuration to undervolt the hardware and intercept any voltage requests coming from users.}
\end{tcolorbox}
\begin{tcolorbox}
    \textit{\textbf{RA2:} a cloud user can uncover such an undervolting strategy by running a simple CPU-bound benchmark until enough processes have failed to render the cloud instance unavailable. The drawback of this detection method is that it is non-selective and cloud instances can fail either soon or late.}
\end{tcolorbox}
\section{Related Work}
\label{sec:rw}

Undervolting the supply voltage for energy savings has been explored 
on CPUs for ARM ~\cite{papadimitriou2017harnessing,papadimitriou2018micro},
x86 processors~\cite{koutsovasilis2020dynamic, papadimitriou2017voltage}
the Itanium micro-architecture~\cite{bacha2013dynamic},
and for \textit{POWER-7} processors~\cite{zu2015adaptive}. 
This experimental
undervolting approach has been extended to GPUs~\cite{leng2015safe} and FPGAs~\cite{DSN20}
as well. 
On the CPU side, frameworks to automate and optimize the process 
of undervolting have been developed~\cite{parasyris2018framework, papadimitriou2017harnessing}.
Recently, \textit{AMD} has announced an undervolting product/framework for their 
most recent \textit{Ryzen} 5000 CPUs~\cite{AMDundervolt}. In~\cite{CCGRID19} the authors discuss the trade-off
between the reduced energy cost and the SLA violation penalties introduced by higher node failures of undervolted
\textit{X86} and \textit{ARM} nodes. 
In CLKSCREW~\cite{tang2017clkscrew}, the undervolting capabilities of 
modern ARM processors is exploited to compromise system security, by targetting undervolting 
faults to specific hardware components to extract cryptographic keys.
\section{Conclusion and Open Challenges}
\label{sec:conc}

A cloud provider can obfuscate the undervolting of processors and even run workloads up to \SI{37}{\percent} more energy-efficiently.
However, by undervolting its infrastructure, the cloud provider incurs a major risk.
Not only does the cloud provider reduce the margin of error but also the system's stability is at stake.
Cloud users can with high probability detect such situations and exploit them using a simple CPU-bound benchmark. %
To some extent, the cloud provider can mitigate stability issues with appropriate cooling systems.
However, it is questionable if the gains of undervolting the infrastructure outweigh the costs of such cooling systems.

Cloud users' options to detect an undervolted ARM instance remain limited and, as shown in this paper, essentially depend on the probability to inject faults non-selectively in processes. %
As our temperature-based guardband analysis and failure evaluation have shown, the higher the processor's temperature, the more likely faults can be injected into processes.
Despite such a powerful cloud provider attacker model, cloud users have an exploitable weak link.
Their only option for presuming a potentially undervolted instance is by increasing the processor's heat dissipation.
Heat dissipation is increased by tuning the CPU frequency and load to the processor's limit.
Under these thermal conditions and an undervolted setup the fault injection probability in processes is rising.
Ideally cloud instances will become unavailable and violate the SLA as a result of continuously failing processes.
Our detection method depends strongly on hardware and how systems such as firmware and AVS react to excessive heat dissipation.
As future plans, we intend to expand this study to a more diverse set of ARM-based hardware targets, focusing in particular on current and future cloud offerings.
We would also like to make our detection method more deterministic by injecting faults in processes more selectively.

\section*{Acknowledgments \& Disclaimer}
We thank the anonymous reviewers for their insightful comments and suggestions.
The views and opinions of the authors do not necessarily reflect those of the U.S. government or Lawrence Livermore National Security, LLC neither of whom nor any of their employees make any endorsements, express or implied warranties or representations or assume any legal liability or responsibility for the accuracy, completeness, or usefulness of the information contained herein.
This  work  was  partially prepared  by  LLNL  under  Contract \texttt{DE-AC52-07NA27344 (LLNL-CONF-817551)} and by the European Union's Horizon 2020 research and innovation programme under the LEGaTO Project (\href{https://legato-project.eu/}{legato-project.eu}), grant agreement No~780681.

{\small   
\bibliographystyle{IEEEtranN}
\bibliography{biblio}
}
\end{document}